\documentstyle[12pt]{article}
\def\simlt{\stackrel{<}{{}_\sim}}
\def\simgt{\stackrel{>}{{}_\sim}}
\def\be{\begin{equation}}
\def\ee{\end{equation}}
\def\bear{\be\begin{array}}
\def\eear{\end{array}\ee}
\def\bea{\begin{eqnarray}}
\def\eea{\end{eqnarray}}

\def\baselinestretch{1}
\begin{document}
\catcode`@=11
\newtoks\@stequation
\def\subequations{\refstepcounter{equation}%
\edef\@savedequation{\the\c@equation}%
  \@stequation=\expandafter{\theequation}
  \edef\@savedtheequation{\the\@stequation}
  \edef\oldtheequation{\theequation}%
  \setcounter{equation}{0}%
  \def\theequation{\oldtheequation\alph{equation}}}
\def\endsubequations{\setcounter{equation}{\@savedequation}%
  \@stequation=\expandafter{\@savedtheequation}%
  \edef\theequation{\the\@stequation}\global\@ignoretrue

\noindent}
\catcode`@=12
\begin{titlepage}

\title{{\bf Some Implications of Charge and Color Breaking in the MSSM}
\thanks{Research supported in part by: the CICYT, under
contracts AEN95-0195 (JAC) and AEN93-0673 (ALL, CM); the European Union,
under contracts CHRX-CT92-0004 (JAC), CHRX-CT93-0132 (CM) and
SC1-CT92-0792 (CM); the
Ministerio de Educaci\'on y Ciencia, under FPI grant (ALL).}
}
\author{ {\bf J.A. Casas\thanks{On leave of absence from Instituto de
Estructura de la Materia CSIC, Serrano 123, 28006 Madrid, Spain.}
${}^{ {\footnotesize,\S}}$},
{\bf A. Lleyda${}^{ {\footnotesize\P}}$ }
and {\bf C. Mu\~noz${}^{\footnotesize\P}$}\\
\hspace{3cm}\\
${}^{\footnotesize\S}$ {\small Santa Cruz Institute for Particle Physics}\\
{\small University of California, Santa Cruz, CA 95064, USA}\\
{\small casas@cc.csic.es}\\
\vspace{-0.3cm}\\
${}^{\footnotesize\P}$ {\small Departamento de F\'{\i}sica
Te\'orica C--XI} \\
{\small Universidad Aut\'onoma de Madrid, 28049 Madrid, Spain}\\
{\small amanda@delta.ft.uam.es $$ cmunoz@ccuam3.sdi.uam.es}}
\date{}
\maketitle
\def\baselinestretch{1.15}
\begin{abstract}
\noindent
We examine some physically relevant implications
of potentially dangerous charge and color breaking minima
for supersymmetric models. 
First, we analyze the stability of the corresponding 
constraints with respect to variations of the 
initial scale for the running of the soft breaking terms, finding that
the larger the scale is, the stronger the bounds become.
In particular, by taking
$M_P$ rather than $M_X$ for the initial scale, which is a more
sensible election, we find substantially stronger and very important 
constraints.
Second, we find general bounds on the universal gaugino mass, the universal 
scalar mass, $m\geq 55$ GeV, and the Higgs bilinear coefficient,
$|B|\simlt 3m$.
Finally, we study the infrared fixed point solution of the top quark mass.
Again, the constraints on the parameter space turn out to be very 
important, including the analytically derivable
bound $\left|M/m\right|\simlt 1$.

\end{abstract}

\thispagestyle{empty}

\leftline{}
\leftline{FTUAM 96/20}
\leftline{SCIPP-96-21}
\leftline{IEM-FT-130/96}
\leftline{May 1996}
\leftline{}

\vskip-22.8cm
\rightline{}
\rightline{ FTUAM 96/20}
\rightline{SCIPP-96-21}
\rightline{ IEM-FT-130/96}
\vskip3in

\end{titlepage}
\newpage
\setcounter{page}{1}

\section{Introduction}

Recently there has been some activity in trying to constrain the soft
parameter space of supersymmetric (SUSY) models [1--7] through the possible
existence of dangerous charge and color breaking minima [8--11].
In ref.\cite{CCB} a systematic discussion of all the potentially dangerous
directions in the field space of the minimal supersymmetric standard model
(MSSM) was carried out.
Imposing that the SUSY standard vacuum should be deeper than the charge and
color  breaking minima, the corresponding constraints are very strong.
Important bounds, not only on the value of the trilinear scalar term ($A$),
but also on the values of the bilinear scalar term ($B$) and the scalar 
and gaugino masses ($m,M$ respectively), are produced on these 
grounds\footnote{Very strong bounds can also be obtained in 
a SO(10) GUT \cite{Strumia}. In the context of String Theory, in particular
when SUSY is broken by the dilaton, the whole parameter space is essentially
excluded \cite{DL}.}.

In the present paper we discuss some issues related to this kind of 
restrictions, which are relevant for supersymmetric model-building,
since they offer a more precise realization of the constraints
under very common circumstances.

The analysis of ref.\cite{CCB} was performed assuming universality of the soft 
breaking terms at the unification scale, $M_X$, as it is usually done in 
the MSSM literature. However, in the standard supergravity (SUGRA) framework,
where SUSY is broken in a ``hidden'' sector, the natural initial scale 
to implement the boundary conditions for the soft terms is 
$M_P\equiv M_{Planck}/\sqrt{8\pi}$ rather than $ M_X$ ($M_P$ is the scale
in which the standard SUGRA Lagrangian is written).
Hence, it is natural to wonder how much the standard charge and color
breaking analysis  (and in particular the results of ref.\cite{CCB}) will get 
modified by taking $M_P$ instead of $M_X$ for the initial scale of the
soft terms\footnote{This question was first pointed out in ref.\cite{Pol} 
for a different type of analysis. In particular the authors studied
the modifications on the low-energy predictions of a SUSY GUT when
non-universal corrections to the soft parameters arise from their
evolution between $M_P$ and $M_X$.}.
This is one of the purposes of the present paper which will be carried out
in section 2.
The second one is to obtain explicit bounds on the supersymmetric
parameters. We will show that the charge and color breaking constraints
put important bounds not only on the value of $A$, but also on the 
values of $M$, $B$ and $m$. Concerning the latter, in ref.\cite{CCB} 
was mentioned the fact that in the limiting case $m=0$
the whole parameter space turns out to be excluded. Now, we will be
more precise, obtaining a general lower bound on $m$. These tasks will be 
accomplished in section 3.
Finally, in section 4, we apply the charge and color breaking constraints to a 
particular region of the parameter space of the MSSM, namely that 
corresponding to the infrared fixed point solution for the top quark mass.
The conclusions are left for section 5.

Of course, we are aware of the possibility of living in a metastable vacuum,
provided that its lifetime is longer than the present age of the universe
\cite{Claudson,Riotto,Kusenko}.
Although this possibility poses some cosmological questions,
as discussed in ref.\cite{DL}, it is clear that could help to rescue 
some regions of the parameter space.
This means that the bounds we consider here are the most 
conservative ones (in the sense of safe ones).
Besides, the identification of the dangerous 
charge and color breaking minima is the first necessary step for
the cosmological analysis.
On the other hand, there are many possible cosmological scenarios,
which requires separate analyses. In particular one can consider 
scenarios where the initial conditions are dictated by
thermal effects (see e.g. refs.\cite{Riotto, Kusenko}) or 
inflationary scenarios. The latter
situation may be much more dangerous and
involved due to the large fluctuations of all the scalar fields,
that could be driven in this way to the dangerous minima.

Let us review the constraints associated with the existence of dangerous 
directions in the field space.
As was mentioned above, a complete analysis of this issue,
including in a proper way the radiative corrections to the
scalar potential, was carried out in ref.\cite{CCB}.
The most relevant results obtained there for our present task
are the following.

There are two types of constraints:
the ones arising from directions in the field-space along
which the (tree-level) potential can become unbounded from below (UFB),
and those arising from the existence of charge and color
breaking (CCB) minima in the potential deeper than the
standard minimum.

Concerning the UFB directions (and corresponding constraints),
there are three of them, labelled as UFB-1, UFB-2, UFB-3
in \cite{CCB}. It is worth mentioning here that in general the
unboundedness is only true
at tree-level since radiative corrections eventually raise the potential for
large enough values of the fields, thus developing minima.
Still these minima can be deeper than
the realistic one (i.e. the SUSY standard model vacuum) and thus dangerous.
The UFB-3 direction, which involves
the scalar fields
$\{H_2,\nu_{L_i},e_{L_j},e_{R_j}\}$ with $i \neq j$
and thus leads also to electric charge
breaking, yields the {\it strongest} bound among {\it all}
the UFB and CCB constraints. 
For the sake of later discussions, let us briefly expose
the explicit form of this bound.
By simple analytical minimization it is possible to write the
value of all the relevant fields along the UFB-3 direction in
terms of the $H_2$ one. Then, for any value of $|H_2|<M_{P}$ satisfying
\be
\label{SU6}
|H_2| > \sqrt{ \frac{\mu^2}{4\lambda_{e_j}^2}
+ \frac{4m_{L_i}^2}{g'^2+g_2^2}}-\frac{|\mu|}{2\lambda_{e_j}} \ ,
\ee
the value of the potential along the UFB-3 direction is simply given
by
\be
\label{SU8}
V_{\rm UFB-3}=(m_2^2 -\mu^2+ m_{L_i}^2 )|H_2|^2
+ \frac{|\mu|}{\lambda_{e_j}} ( m_{L_j}^2+m_{e_j}^2+m_{L_i}^2 ) |H_2|
-\frac{2m_{L_i}^4}{g'^2+g_2^2} \ .
\ee
Otherwise
\be
\label{SU9}
V_{\rm UFB-3}= (m_2^2 -\mu^2 ) |H_2|^2
+ \frac{|\mu|} {\lambda_{e_j}} ( m_{L_j}^2+m_{e_j}^2 ) |H_2| + \frac{1}{8}
(g'^2+g_2^2)\left[ |H_2|^2+\frac{|\mu|}{\lambda_{e_j}}|H_2|\right]^2 \ .
\ee
In eqs.(\ref{SU8},\ref{SU9}) $\lambda_{e_j}$ is the leptonic Yukawa
coupling of the $j-$generation and $m_2^2$ is the sum of the $H_2$ squared
soft mass, $m_{H_2}^2$, plus $\mu^2$. Then, the
UFB-3 condition reads
\be
\label{SU7}
V_{\rm UFB-3}(Q=\hat Q) > V_{\rm real \; min} \ ,
\ee
where $V_{\rm real \; min}=-\frac{1}{8}\left(g'^2 + g_2^2\right)
\left(v_2^2-v_1^2\right)^2$, with $v_{1,2}$ the VEVs of the Higgses $H_{1,2}$,
is the realistic minimum evaluated at $M_S$ (see below)
and the $\hat Q$ scale is given by \linebreak
$\hat Q\sim {\rm Max}(g_2 |e|, \lambda_{top} |H_2|,
g_2 |H_2|, g_2 |L_i|, M_S)$
with
$|e|$=$\sqrt{|H_2||\mu| / \lambda_{e_j} 
}$ and
$|L_i|^2$=$|H_2|^2+|e|^2-\frac{4m_{L_i}^2}{g'^2+g_2^2}$.
Finally, $M_S$ is the typical scale of SUSY masses (normally a good
choice for $M_S$ is an average of the stop masses, for more details
see refs.\cite{Gamberini, bea, CCB}).
Notice from (\ref{SU8},\ref{SU9}) that the negative contribution to $V_{UFB-3}$
is essentially given by the $m_2^2-\mu^2$ term, which can be very sizeable in 
many instances. On the other hand, the positive contribution is dominated by 
the term $\propto 1/\lambda_{e_j}$, thus the larger
$\lambda_{e_j}$ the more restrictive
the constraint becomes. Consequently, the optimum choice for
the $e$--type slepton in eqs.(\ref{SU8}--\ref{SU7}) is the third 
generation one, i.e. ${e_j}=$ stau.

Concerning the CCB constraints, let us mention that the ``traditional'' CCB
bounds \cite{Frere}, when correctly evaluated (i.e. including the
radiative corrections in a proper way), turn out to be extremely weak.
However, the ``improved" set of analytic constraints obtained in
ref.\cite{CCB}, which represent the
necessary and sufficient conditions to avoid dangerous CCB minima,
is much stronger. It is not possible to give here an account of the
explicit form of the CCB constraints used in the present paper. This
can be found in section 5 of ref.\cite{CCB}, to which we refer the
interested reader.

\section{CCB and UFB Constraints and the Initial Scale}

As mentioned in the introduction, the initial boundary conditions for 
the running
MSSM soft terms are usually understood at a scale $M_X$. As we will
see now, bigger initial scales, 
as for example $M_P$, will imply stronger charge and color breaking 
constraints. This can be understood, for example,
from our discussion about the UFB-3 direction above: 
the larger the initial scale 
for the running is, the more important the negative contribution $m_2^2-\mu^2$
to the potential (see eqs.(\ref{SU8}, \ref{SU9})) becomes. 

On the other hand, as discussed in the introduction,
the natural initial scale for the soft SUSY-breaking terms triggered
by the spontaneous breaking of SUGRA is $M_P$.
In Fig.~1 we show the UFB and CCB constraints using $M_P$ as 
the initial scale. More precisely, we plot the 
case\footnote{This value of $B$ is particularly interesting since
it is obtained in several SUGRA theories. See e.g. \cite{CCB,DL} and
references therein.} $B=2m$, with $m=100$, $300$ and $500$ GeV, 
but similar conclusions will be obtained for other
values.
We take $M_{top}^{phys}=174$ GeV as the physical (pole) top mass.
As a matter of fact, 
it is not always possible to choose the boundary condition of the
top Yukawa coupling $\lambda_{top}$ so that the physical (pole) mass is
reproduced because the renormalization group (RG) infrared fixed point of
$\lambda_{top}$ puts an
upper bound on the running top mass $M_{top}$, namely
$M_{top}\simlt 197 sin\beta$
GeV \cite{Inoue}, where $tan \beta$=$ v_2/v_1$.
The corresponding restriction in the parameter space (black region in Fig.~1)
is certainly substantial. 
The region excluded by the CCB bounds is denoted in the figure by circles.
The restrictions coming from the UFB constraints (small filled squares)
are very strong in all the cases.
Most of the parameter space is in fact excluded by the UFB-3
constraint.
Finally, we have also plotted in Fig.1 the region excluded
by the experimental bounds on SUSY particle masses (filled diamonds).
Quite conservatively, we have imposed
\bea
\label{experimentalb}
& &M_{\tilde g} \geq 120\ {\rm GeV}        \;,\;\;
   M_{\tilde \chi^{\pm}}\geq 45\ {\rm GeV} \;,\;\;
   M_{\tilde \chi^o} \geq 18\ {\rm GeV}    \;,\nonumber \\
& &M_{\tilde q}\geq 100\ {\rm GeV}  \;,\;\;\;
   M_{\tilde t} \; \geq 45\ {\rm GeV}\;,\;\;\; \;
   M_{\tilde l} \; \geq 45\ {\rm GeV}\; ,
\eea
in an obvious notation.
The ants indicate regions which are excluded by negative squared mass
eigenvalues.
Notice from Fig.1 that there are areas that are simultaneously constrained
by different types of bounds.
At the end of the day, the allowed region left  (white) is quite small.

This result is to be compared with the one of ref.\cite{CCB}, where the 
assumption of universal soft terms at $M_X$ was taken.
{}From Fig.~3 of that paper we see that the constraints are now substantially
increased and in fact regions of large $M$ which were previously 
allowed become now completely excluded (see Fig.~1 of this paper).
Note also that the UFB bounds are the most sensible ones to 
modifications of the initial scale (for the reasons commented at the
beginning of this section), while the CCB bounds are almost insensitive
to them.

\section{Bounds on MSSM parameters}

{}From Fig.~1 we see that for a given value of $m$ the values of $A$
and $M$ are both bounded from below and above in a correlated way.
E.g. for $-M=m=500$ GeV, we get $1\leq A/m\leq 3.5$; while for
$A/2=m=500$ GeV, we get $-1.75\leq M/m\leq 1$. This restriction
of $M/m$ to a finite and rather narrow range is a novel 
fact\footnote{This has obvious implications for 
gaugino dominance scenarios
(see e.g. ref.\cite{gaudom}), where a large ratio $M^2/m^2$ is found to be
useful in order to solve the FCNC problem.}.

We will show now that the CCB and UFB constraints
put important bounds not only on the value of $A$ and $M$, but also on the 
values of $B$ and $m$, which is also an interesting novel fact.

In Fig.~2 we generalize the previous analysis by varying the value
of $B$ for different values of $m$, namely $m=100$ GeV,
$m=300$ GeV.
The final allowed regions from all types of bounds
in the parameter space of the
MSSM are shown.
Both figures exhibit a similar trend. For a particular value of $m$,
the larger the value of
$B/m$ is, the smaller the allowed region becomes. More precisely, the maximum
allowed value of $B$ for $m=100$ GeV is $B=2.8\ m$, while for 
$m=300$ GeV is $B=2.9\ m$. 
For negative values of $B$
the corresponding figures can easily be deduced from the previous ones,
taking into account that they are invariant
under  the transformation
$B,A,M \rightarrow -B,-A,-M$.
In general, for $m\simlt 500$ GeV, $B$ has 
to satisfy the bound
\be
\label{Bbound}
|B|\simlt 3\ m\;\;.
\ee
This behaviour comes mainly
from the enhancement of the forbidden areas by the UFB-3 constraint and the
requirement of $M^{\rm phys}_{\rm top}=174$ GeV.
Both facts are 
due to the decreasing of $\tan \beta$ as the low-energy value
of $B$ grows.
Then higher top Yukawa couplings are needed in order to reproduce
the experimental top mass. On the one hand, this cannot be always
accomplished due to the infrared fixed point limit on the top
mass. On the other hand, the larger the top Yukawa coupling is, the
stronger the UFB-3 bound becomes.
The same two effects are obtained when $M^{\rm phys}_{\rm top}$ 
is increased and therefore, the
larger the top mass is, the stronger the constraints become.

On the other hand, figures 1, 2 show a clear trend in the sense that 
the smaller the value
of the soft scalar masses, $m$, the more restrictive the constraints become.
This is mainly due to the effect of the UFB-3 constraint (note the almost
exact $m$ invariance of the CCB bounds).

In fact, doing the same type of analysis as in Fig.~2 it is possible to 
find a value of $m$ for which the whole parameter space turns out to be 
excluded. This interesting lower bound on $m$ is
\be
\label{mbound}
m\geq 55\ {\rm GeV}\;\;.
\ee
{}From the above discussion it is evident that the limiting case $m=0$ 
is also excluded. Of course, this has obvious implications for no-scale
models \cite{Lah}. As an example we have plotted in Fig.~3 the case $B=A$.

{}From all the figures, it is clear that the CCB and UFB bounds put also
important constraints on $A$ (as has been traditionally considered)
and $M$, although the precise form of these constraints depends on
the values of the other MSSM parameters and cannot be cast in 
simple general formulae as eqs.(\ref{Bbound}, \ref{mbound}).

Finally, let us mention that the previous bounds would be weaker
if one uses $M_X$ as initial scale (see ref.\cite{CCB}). In particular,
eq.(\ref{mbound}) would then change to
$m\geq 50\ {\rm GeV}$.

\section{Infrared Fixed Point Model}

Let us now apply the UFB and CCB constraints to a particular and attractive
MSSM scenario, namely when the top Yukawa coupling at high energy is large
enough to be in the infrared 
fixed point regime\footnote{For a review 
see e.g. \cite{Waca} and references therein.}. Then 
the running top mass is approximately given by $M_{top}\simeq 197 
sin\beta$ and therefore $tan\beta$ is fixed once a particular value for 
$M^{\rm phys}_{\rm top}$ is chosen 
(in our case $M^{\rm phys}_{\rm top}=174$ GeV).
Thus the number of independent soft parameters is reduced since for given
values of $m$, $M$ and $A$, the value of $B$ is fixed through $tan\beta$.
In addition, the value of $A_t$ at low energies is also fixed in an 
infrared fixed point value (this is not the case for the other trilinear
couplings), and therefore is not an independent parameter
for the electroweak breaking process. Consequently, the required value
of $\mu$ in order to get the correct amount of breaking is determined
only by the values of $m$ and $M$. 

As a consequence of the previous characteristics, it is easy to check that
the UFB-3 constraint, eqs.(\ref{SU6}--\ref{SU7}), does not depend on
the value of $A$ in the strict fixed point limit, while the CCB bounds do.

In Fig.~4 we show the UFB and CCB constraints for the cases $m$=100 and 500
GeV. The figure corresponds to  positive sign of $\mu$
which corresponds to negative value of $B$ at low energy. The results
for negative  sign of $\mu$ can be deduced by taking into account
the invariance of all the results under the transformation
$B,A,M \rightarrow -B,-A,-M$. 

Notice the almost exact independence of the UFB constraints on the
value of $A$, as announced. Actually, the slight departure from this 
independence is mainly due to the 
contribution of the UFB-1 and UFB-2 constraints
in addition to the UFB-3 one. It is clear from the figure that the UFB-3
constraint is very well approximated by 
\be
\label{Mm}
\left|M/m\right|\simlt 1\;\;. 
\ee
It is interesting to note that an approximate form of this bound can
be obtained in an analytical way as follows.

Although the UFB-3 bound must be satisfied for any possible value
of $H_2$ (see eqs.(\ref{SU6}--\ref{SU7})) a weaker bound arises by
fixing $H_2$ (and thus $\hat Q\simeq |H_2|$) at a convenient value. 
To do this, notice from 
eq.(\ref{SU8}) that since $m_2^2 + m_{L_i}^2>0$,
$V_{\rm UFB-3}$ can only take negative values
for $|H_2|\simgt O(10^2)\times m_L^2/\mu$. A particular simple form
for the bound is obtained for a large enough value of $|H_2|$.
For example, taking $|H_2|= 10^6$ GeV (which corresponds
to $t=\log(M_P^2/|H_2|^2)=57$) $V_{\rm UFB-3}$ becomes
\be
\label{VUFB3}
V_{UFB-3}\simeq m^2|H_2|\left\{
0.5\times 10^6\left(1-0.67x^2\right) + 3\left|\frac{\mu}{\lambda_\tau}\right|
\left(1+0.32x^2\right)\right\}\;,
\ee
where we have applied the RG equations of $m_{H_2}^2=m_2^2-\mu^2$, 
$m_L^2$ and $m_e^2$ to write their values at $\hat Q$, and 
$x=M/m$. Taking into account that for most of the parameter space
$|\mu/\lambda_\tau|\ll 10^6$ GeV,
the second term within brackets in eq.(\ref{VUFB3}) is negligible and 
the UFB-3 bound becomes simply $x^2=(M/m)^2<1/0.67=1.49$, i.e. $|M/m|<1.22$.

A more refined bound comes by carefully adjusting  
$|H_2|$ and $\hat Q$ to lower values. However, then 
the bound becomes slightly dependent on the value of $\mu$, and thus
on the value of $m$ since both are correlated, as it is apparent
in Fig.4.


%
%

Concerning the experimental bounds on SUSY particle masses, let us mention
that the region excluded in Fig.~4 is mainly due to charginos and 
neutralinos. Since their masses are independent of the value of $A$,
so the corresponding experimental constraints are.
At the end of the day, the allowed region (white) left is quite small.

Finally, let us point out that Fig.~4 was obtained imposing 
the soft terms boundary conditions at the most sensible (Planck) scale.
Had we
chosen $M_X$ instead of $M_P$, the constraints would have been
less strong. For example, for $m=500$ GeV the empirical form
of the UFB-3 bound would be  $\left|M/m\right|\simlt 1.3$  
instead of $\left|M/m\right|\simlt 1$.

\section{Conclusions}

In this paper we have analyzed some physically relevant implications
of the possible existence of dangerous charge and color breaking minima
for supersymmetric models. First, we have noted that the strength
of the corresponding (CCB and UFB) constraints is quite sensitive 
to the value of the initial scale for the running of the soft breaking terms. 
The larger the scale is, the stronger the bounds become. 
In particular, by taking
$M_P$ rather than $M_X$ for the initial scale, which is a more
sensible election, we find substantially stronger and very important
constraints.

We have also shown that the CCB and UFB constraints
put important bounds not only on the value of $A$ but also on the
values of $M$, $B$ and $m$, which is an interesting novel fact. More precisely,
the values of $A$
and $M$ are both bounded from below and above in a correlated way and
we get the bounds $|B|\simlt 3m$ and $m\geq 55$ GeV.
The lower bound on $m$ has obvious implications for no-scale models.

Finally, we have studied the interesting case of 
the infrared fixed point solution for the top quark mass.
Again, the constraints on the parameter space turn out to be very 
important, including the analytically derivable
bound $\left|M/m\right|\simlt 1$.




\def\MPL #1 #2 #3 {{\em Mod.~Phys.~Lett.}~{\bf#1}\ (#2) #3 }
\def\NPB #1 #2 #3 {{\em Nucl.~Phys.}~{\bf B#1}\ (#2) #3 }
\def\PLB #1 #2 #3 {{\em Phys.~Lett.}~{\bf B#1}\ (#2) #3 }
\def\PR  #1 #2 #3 {{\em Phys.~Rep.}~{\bf#1}\ (#2) #3 }
\def\PRD #1 #2 #3 {{\em Phys.~Rev.}~{\bf D#1}\ (#2) #3 }
\def\PRL #1 #2 #3 {{\em Phys.~Rev.~Lett.}~{\bf#1}\ (#2) #3 }
\def\PTP #1 #2 #3 {{\em Prog.~Theor.~Phys.}~{\bf#1}\ (#2) #3 }
\def\RMP #1 #2 #3 {{\em Rev.~Mod.~Phys.}~{\bf#1}\ (#2) #3 }
\def\ZPC #1 #2 #3 {{\em Z.~Phys.}~{\bf C#1}\ (#2) #3 }

\newpage

\section*{Figure Captions}

\begin{description}
\item[Fig.~1] Excluded regions in the parameter space of the MSSM with
$B$=$2m$.
The black region is excluded because it is not possible to reproduce the
experimental mass of the top.
The small filled squares indicate regions excluded by UFB constraints.
The circles indicate regions excluded by CCB constraints.
The filled diamonds correspond to regions excluded by the experimental lower
bounds on SUSY-particle masses.
The ants indicate regions excluded by negative scalar squared mass eigenvalues.
\item[Fig.~2]  Contours of allowed regions in the parameter space of the MSSM,
for different
values of $B$ and $m$, by the whole set of constraints.
\item[Fig.~3] The same as Fig.1 but with $m$=$0$, $B$=$A$.
\item[Fig.~4] Excluded regions, with the same conventions than in Fig.~1,
in the parameter space of the infrared fixed point model.

\end{description}


\input{psfig.tex}

\newpage
\thispagestyle{empty}
\psfig{file=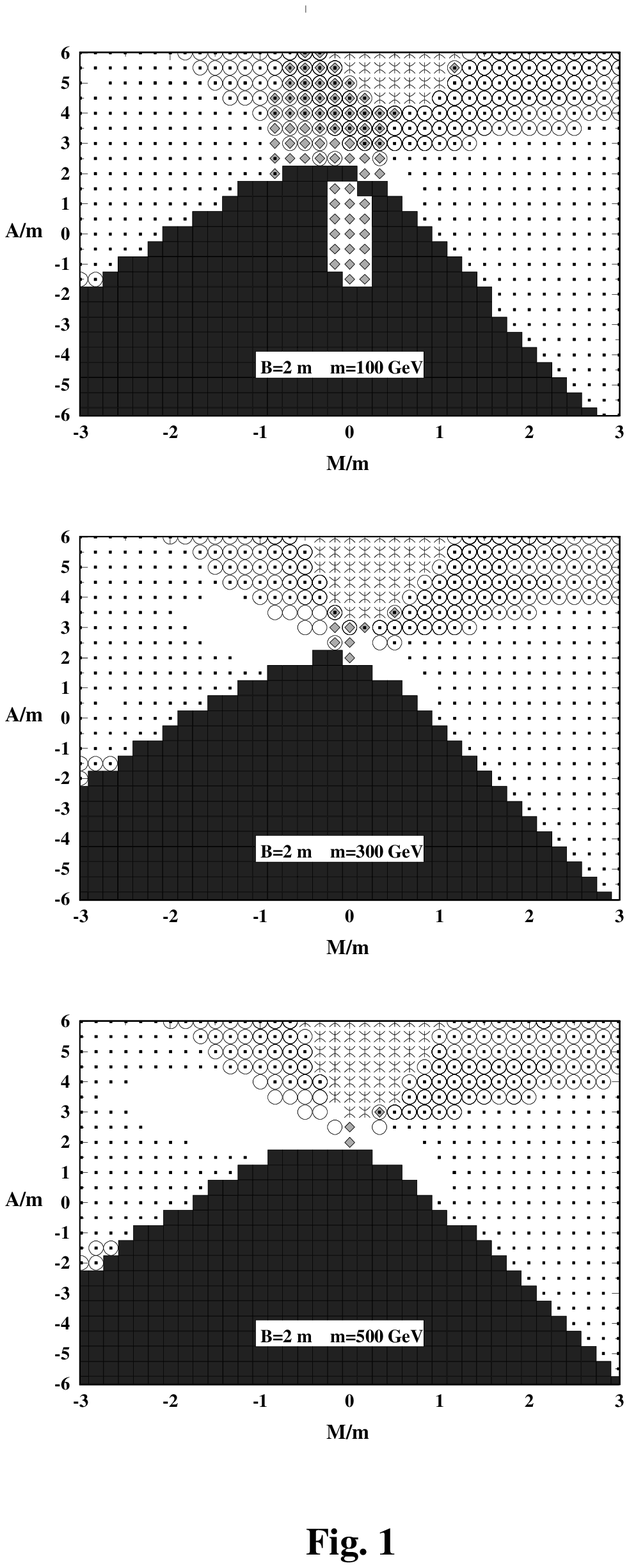,height=23.1cm,angle=180}

\newpage
\thispagestyle{empty}
\psfig{file=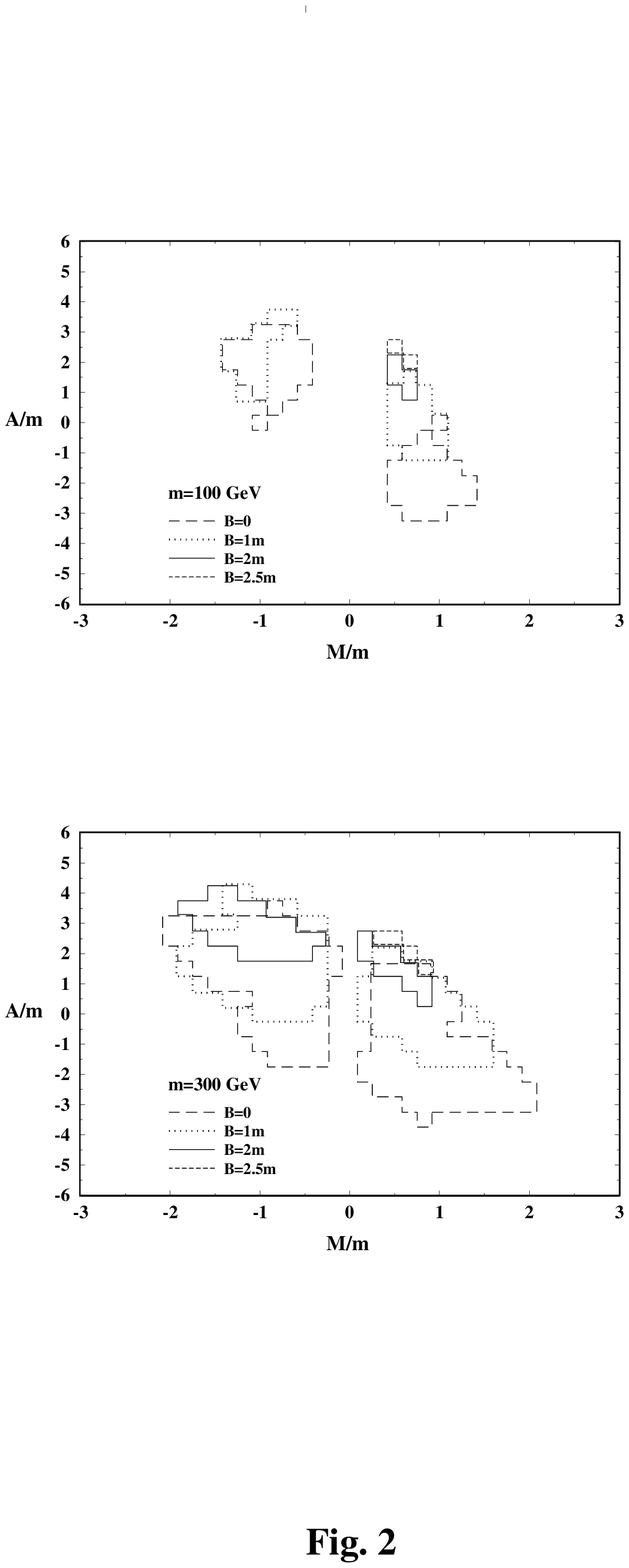,height=23.1cm,angle=180}

\newpage
\thispagestyle{empty}
\psfig{file=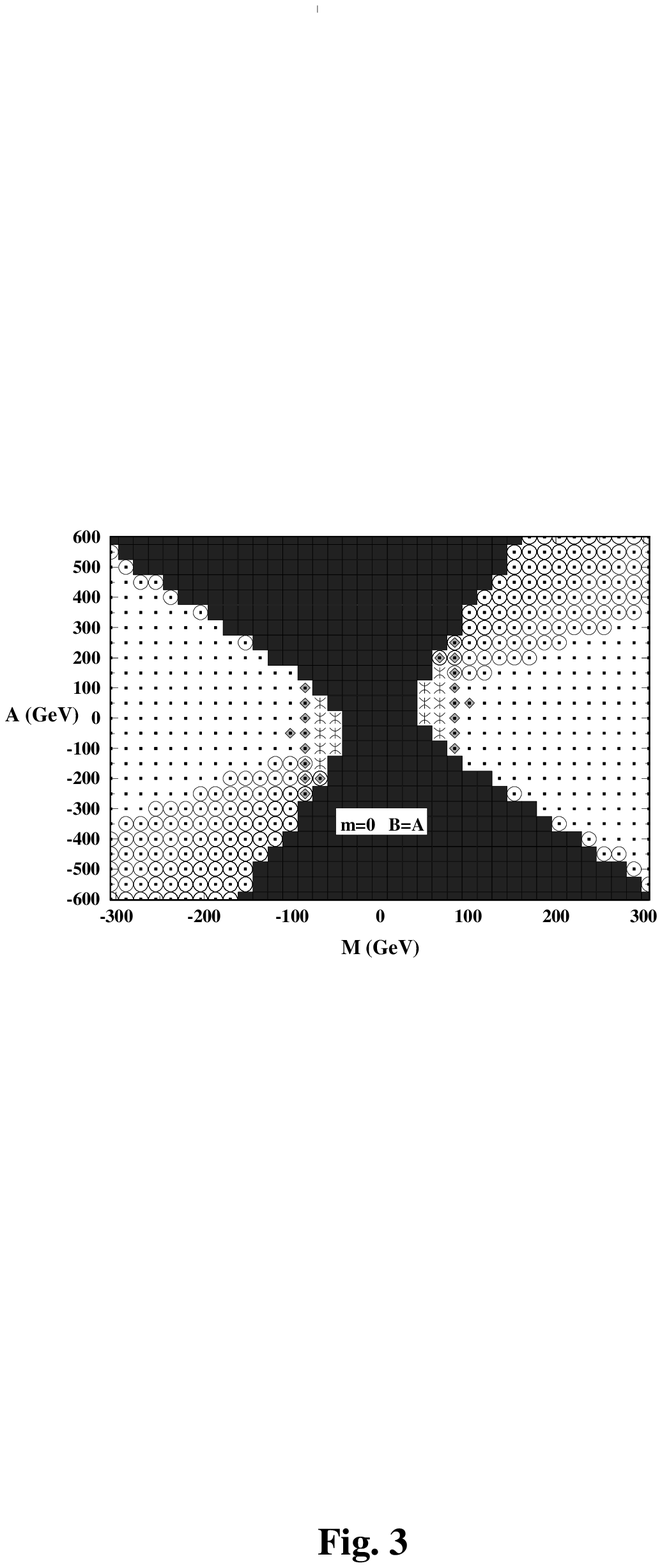,height=23.1cm,angle=180}

\newpage
\thispagestyle{empty}
\psfig{file=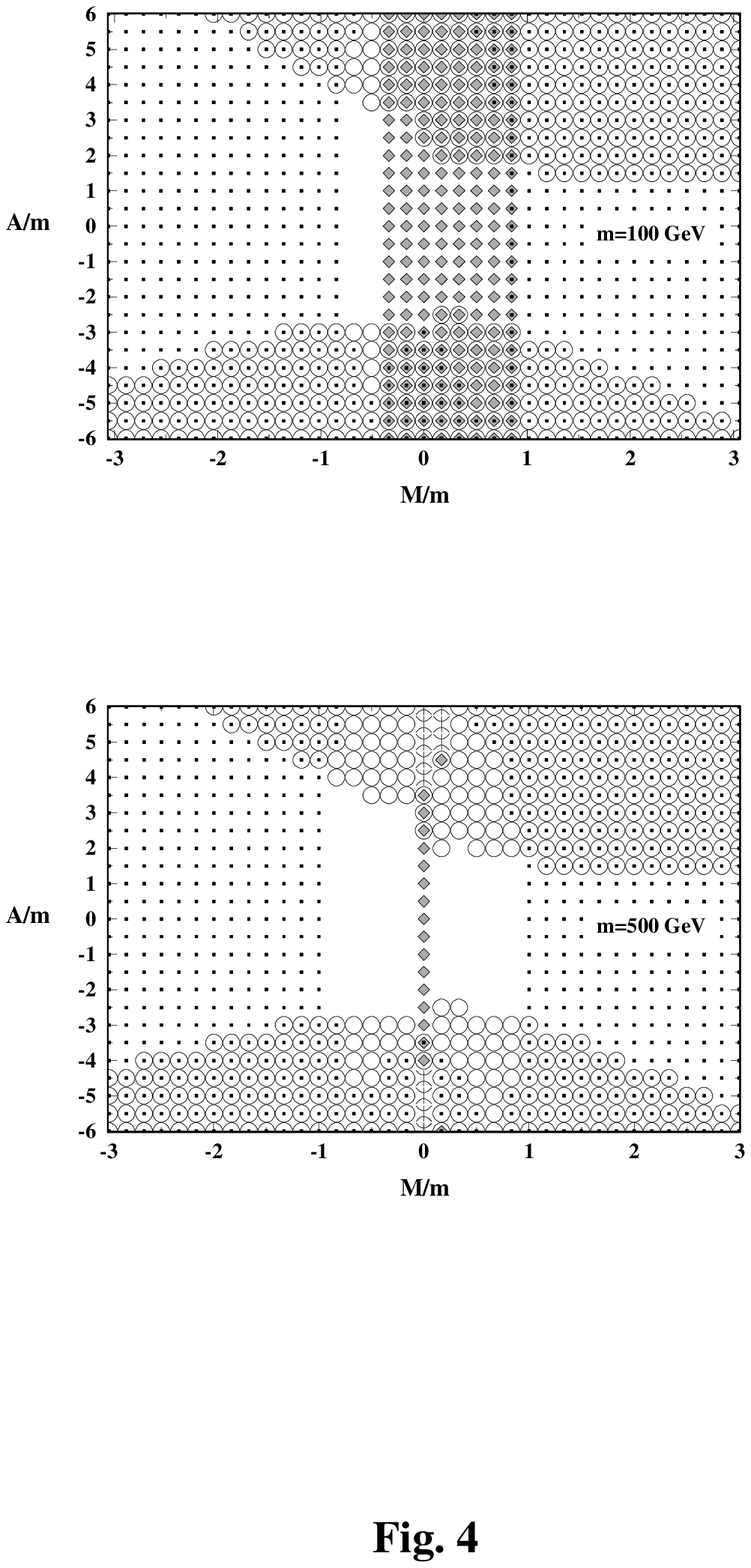,height=23.1cm,angle=180}



\begin{thebibliography}{99}
%
\bibitem{Langacker} P. Langacker and N. Polonsky, \PRD 50 1994 2199; \\
A. Bordner, {\it KUNS-1351}, {\it hep-ph/9506409}.
%
\bibitem{CCB} J.A. Casas, A. Lleyda and C. Mu\~noz, {\it FTUAM 95/11},
{\it hep-ph/9507294}, to appear in \NPB 471 1996 1.
%
\bibitem{Falk} T. Falk, K. Olive, L. Roszkowski and M. Srednicki, 
\PLB 367 1996 183.
%
\bibitem{Riotto} A. Riotto and E. Roulet, {\it SISSA-163/95/EP,
hep-ph/9512401}.
%
\bibitem{DL} J.A. Casas, A. Lleyda and C. Mu\~noz, {\it FTUAM 96/03},
{\it hep-ph/9601357}, to appear in {\it Physics Letters B}.
%
\bibitem{Kusenko} A. Kusenko, P. Langacker and G. Segre, {\it UPR-677-T}, 
{\it hep-ph/9602414}.
%
\bibitem{Strumia} A. Strumia, {\it FTUAM 96/14}, 
{\it hep-ph/9604417}.
%
\bibitem{Frere}J.M. Frere, D.R.T. Jones and S. Raby, \NPB 222 1983 11;\\
L. Alvarez-Gaum\'e, J. Polchinski and M. Wise, \NPB 221 1983 495;\\
J.P. Derendinger and C.A. Savoy, \NPB 237 1984 307;\\
C. Kounnas, A.B. Lahanas, D.V. Nanopoulos and M. Quir\'os, \NPB 236 1984 438.
%
\bibitem{Claudson}M. Claudson, L.J. Hall and I. Hinchliffe, \NPB 228 1983 501.
%
\bibitem{Drees}M. Drees, M. Gl\"uck and K. Grassie, \PLB 157 1985 164;\\
J.F. Gunion, H.E. Haber and M. Sher, \NPB 306 1988 1;\\
H. Komatsu, \PLB 215 1988 323.
%
\bibitem{Gamberini}G. Gamberini, G. Ridolfi and F. Zwirner, \NPB 331 1990 331.
%
\bibitem{Pol} N. Polonsky and A. Pomarol, \PRL 73 1994 2292.
%
\bibitem{bea} B. de Carlos and J.A. Casas, \PLB 309 1993 320.
%
\bibitem{Inoue} K. Inoue, A. Kakuto, H. Komatsu and S. Takeshita,
\PTP 67 1982 1889;\\
L.E. Ib\'{a}\~{n}ez and C. L\'opez, \PLB 126 1983 54; \\
L. Alvarez-Gaum\'e, J. Polchinski and M. Wise, in ref.[7].
%
\bibitem{gaudom} M. Dine, R. Leigh and A. Kagan, \PRD 48 1993 4269;\\
A. Brignole, L.E. Ib\'{a}\~{n}ez and C. Mu\~noz,
\NPB 422 1994 125 [Erratum: {\bf B436} (1995) 747]; \\
 D. Choudhury, F. Eberlein, A. Konig, J. Louis and S. Pokorski,
\PLB 342 1995 180;\\
B. de Carlos, J.A. Casas and J.M. Moreno, \PRD 53 1996 6398.
%
\bibitem{Lah}For a review, see: A.B. Lahanas and D.V. Nanopoulos, 
{\em Phys.~Rep.}~{\bf 145}\ (1987) 1, and references therein.
%
\bibitem{Waca} M. Carena and C.E.M. Wagner, Proceedings of the 2nd
IFT Workshop, Gainesville (1994), {\it hep-ph/9407208}.
%


\end{thebibliography}
\end{document}